\documentclass[reqno]{amsart}
\usepackage{graphics,amsmath}

\def\fs{f} 
\def\rs{s} 

\newcount\n  \newdimen\w

\def\Repeat#1#2{\n=#1\relax\loop\ifnum       
  \n>0\relax #2\advance\n by-1\repeat}

\long\def\OMIT#1{\relax }  

\def\re#1{(\ref{#1})}   
\def\eqn#1#2{ \begin{align} \label{#1}         #2 \end{align}}

\def\nl#1{          \\ \label{#1}        }  
\def\nnl#1{ \tag*{} \\ \label{#1}        }  

\def\delim#1#2#3{\csname\ifcase#1 relax\or   
   big\or Big\or bigg\or Bigg\fi\endcsname   
  {\ifcase#2\or\Delim#3\or\deliM#3\fi}}      
\def\Delim#1{\ifcase#1\relax\or(\or[\or\{\or<\or\langle\or|\or\|\or---{ }\fi}
\def\deliM#1{\ifcase#1\relax\or)\or]\or\}\or>\or\rangle\or|\or\|\or{ }---\fi}

\def\largerfrac#1#2#3{      
  \whichtypesize\n=\currenttypesize\advance\n by #1 \mathchoice
  {\setbox0\hbox{$\displaystyle-$} \w=.5\ht0\advance\w by-.5\dp0\setbox0
    \hbox{\typesize\n $\displaystyle-$} \advance\w by -.5\ht0\advance\w
    by .5\dp0\raise\w \hbox{\typesize\n$\displaystyle{\frac{#2}{#3}}$}}
  {\setbox0\hbox{$-$} \w=.5\ht0 \advance\w by -.5\dp0 \setbox0\hbox
    {\typesize\n $-$} \advance\w by-.5\ht0\advance\w by
    .5\dp0\raise\w\hbox{\typesize\n$\frac{#2}{#3}$}}
  {\setbox0\hbox{$\scriptstyle-$} \w=.5\ht0 \advance\w by-.5\dp0\setbox0
    \hbox{\typesize\n $\scriptstyle-$} \advance\w by -.5\ht0 \advance\w
    by .5\dp0 \raise\w\hbox{\typesize\n$\scriptstyle{\frac{#2}{#3}}$}}
  {\setbox0\hbox{$\scriptscriptstyle-$} \w=.5\ht0
    \advance\w by -.5\dp0 \setbox0\hbox{\typesize\n
    $\scriptscriptstyle-$} \advance\w by -.5\ht0 \advance\w by .5\dp0
    \raise\w\hbox{\typesize\n$\scriptscriptstyle{\frac{#2}{#3}}$}}  }

\begin{document}

\title{Weakly nonlocal nonequilibrium thermodynamics: the Cahn-Hilliard equation}
\author{P. V\'an$^{1,2,3}$ }
\address{$^1$Department of Theoretical Physics, Wigner Research Centre for Physics, H-1525 Budapest, Konkoly Thege Miklós u. 29-33., Hungary; 
$^2$Department of Energy Engineering, Faculty of Mechanical Engineering,  Budapest University of Technology and Economics, 1111 Budapest, Műegyetem rkp. 3., Hungary and 
$^3$ Montavid Thermodynamic Research Group, Budapest, Hungary}
 
\date{\today}

\begin{abstract}
The Cahn-Hilliard and Ginzburg-Landau (Allen-Cahn) equations are derived from the second law. The intuitive approach by separation of full divergences is supported by a more rigorous method, based on Liu procedure and a constitutive entropy flux. Thermodynamic considerations eliminate the necessity of variational techniques and explain the role of functional derivatives. 
\end{abstract}

\maketitle

\section{Introduction}

In continuum theories the extension of classical, well known evolution equations is one of the most exciting theoretical question, where several, completely different approaches compete. When compared to the kinetic theory or nonequilibrium statistical physics the advantage of pure phenomenological ideas is their universality. This way one can analyse the consequences of general requirements, like the basic balances and the second law of thermodynamics without assuming and introducing particular microscopic structures or mechanisms \cite{Gur96a,Gio09a,Van13p2}. In this respect the so called phase-field theories are particularly interesting, because there the influence of microstructure is introduced mostly indirectly, through fluctuating field quantities. The apparent universality of these descriptions is often attributed either to second order phase transitions or background linear instabilities  (see e.g. \cite{HohHal77a,HohKre15a}). On the other hand, from a continuum point of view the obtained macroscopic models are  weakly nonlocal extensions of the simplest evolution equations. For example  the extension of a relaxation dynamics of a single internal variable leads to the Ginzburg-Landau (Allen-Cahn) equation \cite{LanGin50a,LanKha54a,Cah61a,AllCah79a}, or the extension of a diffusion-Fourier dynamics results in the Cahn-Hilliard equation \cite{CahHil58a}. In  phase-field models the role of the second law is rarely constructive, it is used more restrictively. Thermostatics is based on free energy and entropy appears only when thermal phenomena is considered \cite{PenFif90a,AltPaw92a,Ant95a,Ant96a}, or sometimes when dissipation is calculated (e.g. \cite{Grm08a,AndWei11a}). Some more elaborated analyses introduce extra entropy flux \cite{FabEta06a,Gio09a}. The weakly nonlocal extension is due to variational techniques applied together with local equilibrium based thermodynamic considerations \cite{AndWei11a}. In phase-field theories the second law compatibility in the weakly nonlocal sense, for the higher order spatial derivatives is rarely mentioned, and then it is restricted to the Ginzburg-Landau theory. The case of the Cahn-Hilliard equation the large constitutive state space leads to technical difficulties.

However, for more complicated constraints, like the classical balances of continuum mechanics of solids, a simple variational approach may become problematic. The conceptual problem is the doubled theoretical structure. Usually the relaxation dynamics is introduced without a referring to nonequilibrium thermodynamics (e.g. \cite{AndEta98a}). The methodology can be extended and combined to complete and productive theories, like GENERIC \cite{Ott05b,GrmOtt97a,OttGrm97a,Grm08a}, where a clear separation of dissipative and nondissipative parts of the dynamics is based on a bracket formalism with functional derivatives for weak nonlocality. That can keep strict thermodynamic compatibility far beyond the usual phase-field approaches. 

This doubled theoretical structure is due to a lack of a resolution of an interesting and ancient question regarding the origin of evolution equations of physics\footnote{That question goes back at least to the greek philosophers Platon and Aristotle, see and overview in \cite{PriStr95b}.}. Should we separate the ideal world from the real one or is there a common origin of nondissipative and dissipative dynamics? There are two evident strategies to answer this question in a modern context. One either try to extend the variational principles to dissipative dynamics or derive the nondissipative evolution from the second law. The historical roots of the first approach are going back to Helmholtz \cite{YouMan99b} and to a strategy to overcome some strict mathematical conditions \cite{VanMus95a,VanNyi99a}. In order to obtain dissipative evolution and parabolic partial differential equations from variational principles one need to use some inconventional methods \cite{Gya70b,SieFar05b,Ver14a,Gla15a}. However, in most cases the doubled theoretical structure is preserved in a modified form \cite{MatAta05a}. 

The other possibility is pioneered in nonequilibrium thermodynamics, where a deeper mathematical consistency indicated that the second law can be sufficient to derive evolution equations. Anyway, a nondissipative evolution is a special dissipative one, where the dissipation is zero. There exist  an intuitive and a rigorous treatment for weakly nonlocal theories, where the constitutive functions depend on the derivatives of the state variables \cite{Cap85a,Cap89b,Mar02a}. The intuitive technique, the identification of the entropy flux by separation of full divergences, appears independently in many papers and books \cite{CahHil58a,Gro59b,GroMaz62b,Mau99b,Gio09a,HeiEta12a}. The rigorous treatments are the Coleman-Noll and the Liu procedures \cite{ColNol63a,Liu72a}, together with a strict interpretation of the second law \cite{ColMiz67a,MusEhr96a}. This rigorous methods are equivalent \cite{TriAta08a} and interpret the second law as a constrained inequality, where the identification of independent variables and the constitutive state space are the key aspects for handling differential equations as constraints. 

In principle weak nonlocality can be both in time and space, that is the constitutive functions may depend both on time and space derivatives. However, in the following we use the terminology only for the spatial case if not mentioned otherwise. This is the safe approach because we want to avoid spacetime related objectivity questions: spatial derivatives in nonrelativistic space-time are frame independent \cite{Van17a}.

The first applications of the strict mathematical requirements of the second law led to blockingly restrictive results. Coleman and Gurtin has proved that for internal variables only local evolution is possible \cite{ColGur67a}, and Gurtin has shown that in case of pure elasticity there is no weakly nonlocal extension \cite{Gur65a}. These results effectively prevented to understand the universal background of the Ginzburg-Landau (Allen-Cahn) equation and phase-field theories in general. The  thermodynamic investigations of higher grade fluids and solids was slowed down. Additional concepts were necessary to circumvent the restrictive conditions. 

These concepts are very different. For example {\em interstitial work}  generalizes the energy balance \cite{DunSer85a}, the configurational forces generalize the energy and momentum balances \cite{Gur96a,Gur00b}, and the {\em virtual power} is based on a kinematic interpretation of internal variables and a related modification of the energy balance \cite{Ger73a,Mau80a,Fre01b,Mau13a}.\footnote{The phase-fields themselves have their particular interpretation contexts like order parameters, interfacial free energy, diffuse interface, etc ... These all are disguising the general conditions and the universal background.} These interpretational issues elevated further communication barriers, and the pitchforking of thermodynamic theories continued \footnote{This is somehow deeply related to the situation described in the first chapter of \cite{Mau99b}. The specialization is a logical consequence.}. In principle all these extension are capable to construct reasonable conditions and lead to very similar results. 

Here, in this paper we argue that there is a minimal set of assumptions --  it is essentially the second law -- that can generate the Ginzburg-Landau and Cahn-Hilliard type evolution without any further ado. Moreover, the second law can be applied at two levels. The mentioned sophisticated exploitation procedures of Coleman-Noll and Liu can be applied when the blocking barriers are removed: one need to treat the entropy current density as a constitutive quantity and prolong the constraints according to the nonlocality level of the state space. From this point of view the Cahn-Hilliard evolution is particularly instructive, because there the rigorous methods are seemingly hopeless. They are either restricted to check the thermodynamic compatibility of the already derived equations \cite{FabEta06a} or the obtained conditions are not direct and too complicated for practical applications \cite{Paw06a,CimEta16a}. At the same time we will show that the intuitive method of separation of full divergences leads essentially to the same results. This is the classical tool of irreversible thermodynamics \cite{GroMaz62b}, the method used originally by Cahn and Hilliard \cite{CahHil58a} and it is the approach of Gerard Maugin for weakly nonlocal internal variables \cite{Mau99b,Mau06a}. 
Thermodynamic concepts can be used to unite dissipative and nondissipative evolution.

\section{Variational derivation of Ginzburg-Landau and Cahn-Hilliard equations}

In both cases we are looking for the evolution equation of a single scalar internal variable without any constraint in a continuum at rest. In the first case without any additional conditions or constraints for the internal variables, and in the second case with a balance form evolution as a constraint. Here we survey the traditional derivation of both equations which is a characteristic mixture of variational and thermodynamical ideas.

Let us denote \(\xi\) the scalar field. The Helmholtz free energy density, $f$, depends on  this variable and its gradient: $\fs(\xi,\partial_i\xi)$. For the sake of simplicity we assume the following square gradient form, a Ginzburg--Landau free energy function:
\eqn{GL_fun}{
    \fs(\xi, \partial_i\xi)= \fs_0(\xi) + \gamma\partial_i\xi\partial^i\xi/2,
}

\noindent where  $\gamma$ is a nonnegative material parameter, which is scalar for isotropic continua. $\fs_0$ is the classical, local part of the free energy, that may have particular forms, if $\xi$ is an order parameter of a second order phase transition.  $\partial_\xi$ denotes the gradient of $\xi$ and we apply Einstein's summation convention and abstract indices, $i,j,k \in \{1,2,3\}$.\footnote{Please note, these are abstract indices without coordinates \cite{Pen04b}.}. 

Then, following the usual arguments, one assumes, that the rate of \(\xi\)  in a body with volume \(V\) is negatively proportional to the change of the free energy, denoted by $\delta$:
\eqn{int_GL}{
\frac{d}{dt}\int_V \xi {\rm d}V = -l \delta\int_V \fs(\xi,\partial_i\xi)dV.
}

Assuming that this equality is valid for any \(V \) we obtain the general {\em Ginzburg--Landau (Allen-Cahn) equation} in the following form:
\eqn{GL_eq}{
\partial_t\xi = -l   \frac{\delta \fs}{\delta \xi} =
-l \big[\partial_\xi \fs - \partial_i\big(\partial_{\partial_i\xi} \fs\big)\big].
}

Here $\partial_t$ is the partial time derivative, $ \frac{\delta }{\delta \xi}$ in the functional derivative, and $l$ is a material parameter. With the square gradient free energy, \re{GL_fun}, one arrives at the classical form of the equation:
\eqn{sGL_eq}{
\partial_t\xi = 
-l \big[\partial_\xi \fs_0 - \gamma\partial^i_{\ i}\xi\big)\big]. 
}


The second most important basic example of phase-field theories is the Cahn--Hilliard equation, used for modelling phase transitions in solid media. Usually it is introduced as a dynamic equation of a conserved order parameter \cite{HohHal77a}. Therefore now we are looking for a  balance form evolution of an internal variable. Then one may assume a derivation by classical irreversible thermodynamics, where in the thermodynamic force the gradient of the internal variable is substituted by a functional derivative, assuming a nonlocal interaction. The variational origin can be less argumented in this case. Therefore, denoting by $j^i$ the current density of $\xi$, the balance is written as
\eqn{CH_bal}{
\partial_t \xi + \partial_i j^i = 0.
} 

Then, according to classical irreversible thermodynamics the thermodynamic flux is the gradient of the corresponding intensive quantity, $A_\xi$. Without thermal interaction this intensive quantity is the partial derivative of the free energy by $\xi$, that is $A_\xi = \frac{\partial \fs}{\partial \xi}$. Therefore the constitutive equation for the flux is
\eqn{CH_const}{
 j^i = -\kappa \partial^i A_\xi.
}
In case of a weakly nonlocal free energy, $\fs(\xi,\partial_i\xi)$, the partial derivative, is to be substituted by a functional derivative, $\hat A_\xi = \frac{\delta \fs}{\delta \xi}.$ Then we obtain the Cahn-Hilliard equation for the evolution of the internal variable as
\eqn{CH_eq}{
\partial_t \xi -   \partial_i\big(\kappa \partial^i \hat A_\xi\big) = \partial_t \xi -   \partial_i\left(\kappa \partial^i  \big[\partial_\xi \fs - \partial_i\big(\partial_{\partial_i\xi} \fs\big)\big]\right)= 0.
}

These derivations are remarkably simple and sound from a physical point of view also when developed in a more detailed form. On the other hand they combine a variational extremum principle with a frame theory of nonequilibrium thermodynamics. In the next section we will see, that variational considerations are not necessary at all, and they role is simply a separation of surface and bulk contributions for the entropy.

First we will investigate the Ginzburg-Landau equation.

\section{The thermodynamic origin of the Ginzburg-Landau (Allen-Cahn) equation}

\subsection{Separation of full divergences}
 
With this intuitive method one constructs the entropy balance by introducing the constraints through the time derivative of the entropy density and then identify an entropy flux by separating full divergences in the expression. This idea was used independently by in several works \cite{CahHil58a,Gio09a,HeiEta12a}, but as a method was introduced in classical irreversible thermodynamics, where the starting point is the Gibbs relation for the classical extensives \cite{Gro59b,GroMaz62b,Gya70b}. Gerard Maugin applied the approach in several cases, including internal variables \cite{Mau99b,MauDro83a,Mau06a,BerVan17b}. 

Let us assume that the evolution equation of the internal variable $\xi$ is written in a general form as 
\eqn{xievol}{
\partial_t \xi =  F.
}
As we can see one does not fix the domain of the right hand side yet. The starting point is a first order weakly nonlocal entropy density, $s(\xi,\partial_i\xi)$, like free energy density in the previous section. Now we investigate the entropy balance. It is more convenient then with free energy, however, there is no conceptual difference, it is only a question of convenience. The relation of the two approaches and a more systematic background with complete continuum mechanics on material manifolds is developed e.g. in \cite{BerVan17b}. 

Let us calculate the time derivative of the entropy density and separate the full divergences with Leibnitz rule:
\eqn{GL_cal}{
\partial_t \rs(\xi,\partial_i \xi) &=\partial_\xi \rs \partial_t \xi + \partial_{\partial_i\xi} \rs \partial_{ti}\xi = 
- \partial_\xi \rs F - \partial_{\partial_i\xi} \rs \partial_{i} F =        \nnl{a}
&  -\partial_i\left(F \partial_{\partial_i\xi} \rs\right) + F\left(- \partial_\xi\rs + \partial_i(\partial_{\partial_i\xi} \rs )\right).
}

Therefore the entropy flux is identified as
\eqn{Gl_sflu}{
J^i = F \partial_{\partial_i\xi} \rs,
}
and the entropy production is
\eqn{GL_sflu}{
\partial_t \rs + \partial_i J^i = \Sigma =  F\left(- \partial_\xi\rs + \partial_i(\partial_{\partial_i\xi} \rs )\right)  \geq 0.
} 

One can see, that the entropy inequality can be solved with the identification of thermodynamic fluxes and forces and assuming a linear relationship between them
\eqn{GL_epr}{
F = l\left(- \partial_\xi\rs + \partial_i(\partial_{\partial_i\xi} \rs )\right),
} 
where $l>0$ is a material relaxation coefficient. The complete evolution of $\xi$ will be given by an entropic Ginzburg-Landau equation:
\eqn{GLevo1}{
\partial_t \xi =   l\left(- \partial_\xi\rs + \partial_i(\partial_{\partial_i\xi} \rs )\right).
}
One may observe that natural boundary conditions emerge assuming a zero entropy flux. These are convenient for numerical solutions and effectively substitute the natural boundary conditions of variational principles \cite{Gio09a,BerVan17b}.

\subsection{Ginzburg-Landau equation: a more rigorous derivation}

 The method is simple and clear, this is the advantage. The disadvantage is that one cannot fix the constitutive quantities and the constitutive state space in advance, it is determined along the calculations, therefore the results are not unique and require a verification by more rigorous methodology.

In this case one should assume a second order weakly nonlocal state space, spanned by the internal variable field, $\xi$,  and its first and second derivatives,  $\partial_i \xi$ and $\partial_{ij} \xi$.  

We distinguish the
\begin{itemize}
\item {\em space of basic variables}, spanned by $\xi$,
\item the {\em consititutive state space}, spanned by $(\xi,\partial_i \xi, \partial_{ij}\xi)$,
\item and the constitutive functions are $s$, $J^i$ and $F$. 
\end{itemize}

Then the so called {\it process direction space} \cite{MusEhr96a,Van08a} is spanned by the higher derivatives of the constitutive state space, $(\partial_t \xi,\partial_{ti} \xi, \partial_{tij}\xi, \partial_{ijk}\xi)$. We can observe, that now these derivatives are not independent, both the evolution equation, \re{xievol}, and also its derivative 
\eqn{devi1}{
\partial_{ti} \xi + \partial_iF = 0_i,
}
defines a relation in the process direction and constitutive state spaces. Therefore both the evolution equation of $\xi$ and also its gradient are constraints for the entropy inequality. 

In order to apply Liu procedure we introduce the Lagrange-Farkas multipliers, $\lambda$ and $\Lambda^i$, for the equations \re{xievol} and \re{devi1} respectively. Then the application of Liu procedure  leads to
\eqn{GLliu}{
0  & \leq  \partial_t  s+ \partial_i  J^i-
   \lambda\left(\partial_t \xi + F\right)-
   \Lambda^i\left(\partial_{ti} \xi + \partial_iF\right) =
 \nonumber\\
   &=\partial_\xi s\; \partial_t \xi + 
        \partial_{\partial_i\xi} s\; \partial_{it} \xi+
     \partial_{\partial_{ij}\xi} s\; \partial_{ijt} \xi +
      \partial_\xi J^i\; \partial_i \xi + 
        \partial_{\partial_j\xi} J^i\; \partial_{ij} \xi +
        \partial_{\partial_{jk}\xi} J^i\; \partial_{ijk} \xi -
    \nonumber\\
  &\quad -\lambda \left( \partial_t \xi+ F\right) -
     \Lambda^i\left(\partial_{ti} \xi + 
      \partial_\xi F\; \partial_i \xi + 
        \partial_{\partial_j\xi}F\; \partial_{ij} \xi+
     \partial_{\partial_{jk}\xi}F\; \partial_{ijk} \xi  \right) =
     \nonumber\\
  &= \left(\partial_\xi  s - 
        \lambda \right) \underline{\partial_t \xi} +
  \left(\partial_{\partial_i\xi} s - 
        \Lambda^i \right)\; \underline{\partial_{it} \xi}+
    \partial_{\partial_{ij}\xi} s\; \underline{\partial_{ijt} \xi} +
     \nonumber\\
  & \quad+\left( \partial_{\partial_{jk}\xi}J^i- 
         \Lambda^i \partial_{\partial_{jk}\xi}F
                \right)\underline{\partial_{ijk} a}+
        \partial_\xi J^i\; \partial_i \xi + 
        \partial_{\partial_j\xi} J^i\; {\partial_{ij} \xi}-
    \nonumber\\
  & \quad -{\Lambda}^i\left( 
      \partial_\xi f\; \partial_i \xi + 
        \partial_{\partial_j\xi} f\; \partial_{ij} \xi\right)- \lambda F.    
}

The multipliers of the underlined partial derivatives, the members of the process direction space,  give the Liu equations as 
\eqn{liugl1}{
 \partial_t \xi &:&\ \partial_\xi  s &= \lambda, \\
  \partial_{it} \xi &:&\ \partial_{\partial_i\xi}  s &=\Lambda^i, \\
   \partial_{ijt} \xi &:&\ \partial_{\partial_{ij}\xi}  s &=0^{ij}, \nl{liugl2}
  \partial_{ijk} \xi &:&\ \partial_{\partial_{(jk}\xi} J^{i)}&= 
          \Lambda^{(i} \partial_{\partial_{jk)}\xi} f . 
}

The first two equations determine the Lagrange--Farkas multipliers as the derivatives of the entropy, a solution of the third one gives that the entropy is independent on the second gradient of the state variable, \(\xi\). Therefore the Lagrange--Farkas multiplier,  $\Lambda^i$, is also independent of this variable. In the last equation the indexed parenthesis indicate the symmetric part. The last equation can be integrated as
\eqn{GL_sfluf}{
 J^i (\xi,\partial_i \xi, \partial_{ij}\xi)= 
        \partial_{\partial_i\xi}  s(\xi,\partial_i \xi)\ 
         F(\xi,\partial_i \xi, \partial_{ij}\xi) +
        {\mathfrak{J}}^{i}(\xi,\partial_i \xi).
} 

Here the extra entropy current density, $ {\mathfrak{J}}^i$, is an arbitrary constitutive function of the indicated variables and we did not restrict the Liu  condition to the symmetric part of the expression. This is a complete solution of the Liu equations, \re{liugl1}--\re{liugl2}. Considering these results, the dissipation  inequality is reduced to the following form:
\eqn{GL_ep}{
0 \leq \partial_i {\mathfrak{J}}^i +
        \big[\partial_i(  \partial_{\partial_i\xi}  s)-
   \partial_\xi  s\big]   F.
}

Assuming that the extra entropy flux, ${\mathfrak{J}}^i$, is zero, we obtain a product of undetermined constitutive quantity and the derivatives of the entropy, the same  thermodynamic force-flux system as in \re{GL_sflu}. The classical linear solution of the inequality results in 
\eqn{nono}{
 F =  l\; \big[\partial_i(  \partial_{\partial_i\xi}  s)-
   \partial_\xi  s\big], \qquad  l>0.
}
Therefore the evolution equation of an internal variable in a second order weakly nonlocal constitutive state space will be the Ginzburg-Landau equation:
\eqn{fevi11}{
\partial_t \xi =  l\; \big[\partial_\xi  s -
        \partial_i(  \partial_{\partial_i\xi}  s)\big].
}  
This result is a consequence of the second law, independently of any microscopic interpretation. If the entropy density has a square gradient form, \re{GL_fun}, \cite{BedAta03a,JohBed03a,JohBed04a,GlaBed08a}, we obtain the classical form of the equation if \(\gamma\) is constant. The concavity of the entropy requires $\gamma>0$.

\section{The thermodynamic origin of the Cahn--Hilliard equation}
\label{CH-gynl}

\subsection{Separation of full divergences}

In this case the internal variable $\xi$ is conservative, its evolution equation has a balance form
\eqn{CH_bal1}{
\partial_t \xi + \partial_i j^i =0,
}
where $j^i$ is the current density of $\xi$. Now the constitutive functions are the entropy, its current density and the flux of the state variable: $s, J^i$ and $j^i$. When looking at the equation \re{CH_eq} we can observe, that we need at least a fourth order weakly nonlocal state space. In case of the Coleman-Noll or Liu procedures the large number of composite derivatives, and the resulted nonlinearity in the process direction variables, encumbers to find explicit solutions of the inequality. However, the simple separation of the full divergences in the entropy balance gives the expected result.  

Let us assume, that the entropy density depends on the state variable and also on its gradient, $\rs(\xi,\partial_i\xi)$. Let us calculate its time derivative, and substitute the balance, \re{CH_bal1}, as a constraint:
\eqn{CH_cal}{
\partial_t \rs(\xi,\partial_i \xi) &=\partial_\xi \rs \partial_t \xi + \partial_{\partial_i\xi} \rs \partial_{ti}\xi = 
- \partial_\xi \rs \partial_i j^i - \partial_{\partial_i\xi} \rs \partial_{ik} j^k =        \nnl{a1}
&- \partial_i\left(\partial_\xi \rs j^i\right)  + \partial_i\left(\partial_\xi \rs \right)j^i -
	\partial_i\left( \partial_{\partial_i\xi} \rs \partial_{k}j^k\right) + 
	\partial_i\left( \partial_{\partial_i\xi} \rs \right) \partial_{k}j^k =              \nnl{b}
&- \partial_i\left[\partial_\xi \rs j^i + \partial_{\partial_i\xi} \rs \partial_{k}j^k  - 
		\partial_k\left( \partial_{\partial_k\xi} \rs \right) j^i \right]  + 
	\partial_i\left(\partial_\xi \rs  -  \partial_k\left(\partial_{\partial_k\xi} \rs\right) \right) j^i 
}

Therefore the entropy can be identified as
\eqn{CH_sflu}{
J^i =  \left(\partial_\xi \rs - \partial_k\left( \partial_{\partial_k\xi} \rs \right)\right) j^i + \partial_{\partial_i\xi} \rs \partial_{k}j^k,  
}
and the entropy production is
\eqn{CH_epr}{
\Sigma =  
	\partial_i\left(\partial_\xi \rs  -  \partial_k\left(\partial_{\partial_k\xi} \rs\right) \right) j^i  \geq 0.
} 
One can see, that the solution of this inequality is easy with the identification of the thermodynamic force by the gradient of $\partial_i\left(\partial_\xi \rs - \partial_k\left(\partial_{\partial_k\xi} \rs\right)\right)$, and the thermodynamic flux as the current density of the internal variable, $j^i$. For isotropic materials the coefficient is a scalar and we obtain the constitutive equation, \re{CH_const}:
\eqn{CH_constt}{
j^i =  -\kappa
	\partial^i\left(\partial_\xi \rs  -  \partial_k\left(\partial_{\partial_k\xi} \rs\right) \right),
} 
where $\kappa>0$, because of the second law.  Substituting this expression into the balance \re{CH_bal1} we obtain the Cahn-Hilliard equation:
\eqn{CH_balf}{
\partial_t \xi - \partial_i \left[\kappa
	\partial^i\left(\partial_\xi \rs  -  \partial_k\left(\partial_{\partial_k\xi} \rs\right) \right)\right] =0.
}
However, let us recognize that \re{CH_constt} requires that the internal variable flux is a third order weakly nonlocal function, depending on the variables $(\xi,\partial_i \xi, \partial_{ij}\xi, \partial_{ijk}\xi)$.

\subsection{Cahn-Hilliard equation: a more rigorous derivation}

Here we should start from a fourth order weakly nonlocal constitutive state space for a systematic analysis. Therefore, the 
\begin{itemize}
\item space of basic variables is spanned by $\xi$,
\item the consititutive state space is spanned by $(\xi,\partial_i \xi, \partial_{ij}\xi, \partial_{ijk}\xi,\partial_{ijkl}\xi,)$.
\item The constitutive functions are $s$, $J^i$ and $j^i$. 
\end{itemize}

In order to apply the Liu procedure we introduce the Lagrange-Farkas multipliers $\lambda$ for the balance of $\xi$, \re{CH_bal1}, and $\Lambda^j$ for the gradient of \re{CH_bal1}:
\eqn{gCH_bal1}{
\partial_{tj} \xi + \partial_{ij} j^i =0.
}
With a fourth order weakly nonlocal state space one may wonder, whether  higher order spatial derivatives of the constraint should be applied. E.g. Cimmelli argues that the consistent evaluation of the second law in this case requires the third and fourth derivatives as constraints, too \cite{Cim07a}. However, one can prove that in that case the constructive character of the derivation is lost. Therefore, let us calculate the constrained inequality, with  the Lagrange-Farkas multipliers, as in the previous section, but now with different constraints and in a larger state space.
\eqn{CHliu}{
0  & \leq  \partial_t  s+ \partial_i  J^i-
   \lambda\left(\partial_{t} \xi + \partial_{i} j^i\right)-
   \Lambda^j\left(\partial_{tj} \xi + \partial_{ij} j^i\right) =  \nonumber\\
   &=\partial_\xi s\; \partial_t \xi + 
        \partial_{\partial_{i}\xi} s\; \partial_{it} \xi+
	    \partial_{\partial_{ij}\xi} s\; \partial_{ijt} \xi+
	    \partial_{\partial_{ijk}\xi} s\; \partial_{ijkt} \xi+
	    \partial_{\partial_{ijkl}\xi} s\; \partial_{ijklt} \xi+    \nonumber\\
  &\quad +   \partial_{\xi} J^i\; \partial_{i} \xi+
	 \partial_{\partial_{j}\xi} J^i\; \partial_{ij} \xi+
	 \partial_{\partial_{jk}\xi} J^i\; \partial_{ijk} \xi+
	 \partial_{\partial_{jkl}\xi} J^i\; \partial_{ijkl} \xi+
	 \partial_{\partial_{jklm}\xi} J^i\; \underline{\partial_{ijklm} \xi}+   \nonumber\\
 &\quad -\lambda\left( \partial_{t} \xi+   
	    	\left(\partial_{\xi} j^{i}\right)\; \partial_{i} \xi+
   	   		\left(\partial_{\partial_{j}\xi} j^i\right)\; \partial_{ij} \xi+	
			\left(\partial_{\partial_{jk}\xi} j^i\right)\; \partial_{ijk} \xi+\right.\nonumber\\
  &\qquad 	\left. + 	\left(\partial_{\partial_{jkl}\xi} j^i\right)\; \partial_{ijkl} \xi+
			\left(\partial_{\partial_{jklm}\xi} j^i\right)\; \underline{\partial_{ijklm} \xi}\right)\nonumber\\
  &\quad -\Lambda^n \left( \partial_{nt} \xi+   
	    \partial_n\left(\partial_{\xi} j^i\right)\; \partial_{i} \xi+
			\left(\partial_{\xi} j^{i}\right)\; \partial_{in} \xi+
   	    \partial_n\left(\partial_{\partial_{j}\xi} j^i\right)\; \partial_{ij} \xi+
			\left(\partial_{\partial_{j}\xi} j^i\right)\; \partial_{ijn} \xi+\right.\nonumber\\
 &\qquad 	\left.	+   \partial_n\left(\partial_{\partial_{jk}\xi} j^i\right)\; \partial_{ijk} \xi+ 	\left(\partial_{\partial_{jk}\xi} j^i\right)\; \partial_{ijkn} \xi+
	    \partial_n\left(\partial_{\partial_{jkl}\xi} j^i\right)\; \partial_{ijkl} \xi+\right.\nonumber\\
 &\qquad 	\left.	+ 
			\left(\partial_{\partial_{jkl}\xi} j^i\right)\; \underline{\partial_{ijkln} \xi}+
	    \partial_n\left(\partial_{\partial_{jklm}\xi} j^i\right)\; \underline{\partial_{ijklm} \xi} +
			\left(\partial_{\partial_{jklm}\xi} j^i\right)\; \underline{\partial_{ijklmn} \xi}\right)\geq 0   }

The multipliers of the partial time derivatives give the following Liu equations:
\eqn{liuch1}{
 \partial_t \xi &:&\ \partial_\xi  s &= \lambda, \\
  \partial_{it} \xi &:&\ \partial_{\partial_i\xi}  s &=\Lambda^i, \\
   \partial_{ijt} \xi &:&\ \partial_{\partial_{ij}\xi}  s &=0^{ij},\\
   \partial_{ijkt} \xi &:&\ \partial_{\partial_{ijk}\xi}  s &=0^{ijk},\\
   \partial_{ijlkt} \xi &:&\ \partial_{\partial_{ijkl}\xi}  s &=0^{ijkl},
}
The first two equations give the Lagrange--Farkas multipliers as the derivatives of the entropy, and a solution of the last three ones result in that the entropy must depend only on the state variable and its gradient. Therefore the $\lambda$ and $\Lambda^i$  Lagrange--Farkas multipliers depend only on these variables, too. In summary the entropy density and its derivatives are $s= s(\xi,\partial_i\xi)$, $\partial_\xi s= \lambda$, and  $\partial_{\partial_i\xi}s= \Lambda^i$. 

Let us investigate now the multipliers of the underlined sixth and fifth spatial derivatives, the remaining members of the process direction space in the inequality:
\eqn{liuch3}{
   \partial_{ijklmn} \xi &: \partial_{\partial_{i}\xi}  s \partial_{\partial_{jklm}\xi}  j^n =0^{ijklmn},\\
   \partial_{ijklm} \xi &: \partial_{\partial_{ijkl}\xi} J^m =\partial_\xi s  \partial_{\partial_{ijkl}\xi} j^m + 
		\partial_{\partial_{n}\xi}  s\left(\partial_n\left[ \partial_{\partial_{ijkl}\xi} j^m\right] + \delta^i_{n} \partial_{\partial_{jkl}\xi} j^m\right)
}
If the entropy is first order weakly nonlocal, then a solution of the first equation is 
\eqn{c1}{
\ \partial_{\partial_{ijkl}\xi} \left(\partial_n j^i\right) =0.
}
The solution of the second equation will be the following expression for the entropy flux
\eqn{jch}{
 J^i =  \partial_{\xi} s j^i + \partial_{\partial_n\xi} s \partial_n j^i+
       {\mathfrak{J}}^{i}(\xi,\partial_i \xi,\partial_{ij}\xi,\partial_{ijk}\xi),
} 
where the last term, the extra entropy flux, $ {\mathfrak{J}}^i$, is only third order weakly nonlocal. 

In this solution we have used the first of the indenties below:
\eqn{iden}{
   \partial_{\partial_{ijkl}\xi} \left[\partial_n j^m\right] &= 
		\partial_n\left[ \partial_{\partial_{ijkl}\xi} j^m\right] + 
		 \delta^i_{n} \partial_{\partial_{jkl}\xi} j^m,  \nonumber\\
   \partial_{\partial_{ikl}\xi} \left[\partial_n j^m\right] &= 
		\partial_n\left[ \partial_{\partial_{ikl}\xi} j^m\right] + 
		 \delta^i_{n} \partial_{\partial_{kl}\xi} j^m, \nonumber\\
   \partial_{\partial_{il}\xi} \left[\partial_n j^m\right] &= 
		\partial_n\left[ \partial_{\partial_{il}\xi} j^m\right] + 
		 \delta^i_{n} \partial_{\partial_{l}\xi} j^m,  	 \nonumber\\
   \partial_{\partial_{i}\xi} \left[\partial_n j^m\right] &= 
		\partial_n\left[ \partial_{\partial_{i}\xi} j^m\right] + 
		 \delta^i_{n} \partial_{\xi} j^m.  	
}

Using also the other identities of \re{iden} and also the entropy flux, \re{jch}, the dissipation  inequality reduces to the following simple form:
\eqn{chdineq}{
\partial_{ijkl}\xi \left(\partial_{\partial_{jkl}\xi} J^i - 
	\partial_\xi s \partial_{\partial_{jkl}\xi} j^i - 
	\partial_{\partial_{m}\xi}s \partial_{\partial_{jkl}\xi} (\partial_m j^i)\right) +  \nonumber\\
\partial_{ijk}\xi \left(\partial_{\partial_{jk}\xi} J^i - 
	\partial_\xi s \partial_{\partial_{jk}\xi} j^i - 
	\partial_{\partial_{m}\xi}s\partial_{\partial_{jk}\xi}(\partial_m  j^i)\right)+ \nonumber\\
\partial_{ij}\xi \left(\partial_{\partial_{j}\xi} J^i -
	\partial_\xi s \partial_{\partial_{j}\xi} j^i - 
	\partial_{\partial_{m}\xi}s \partial_{\partial_{j}\xi}(\partial_m  j^i)\right) +  \nonumber\\
\partial_{i}\xi \left(\partial_{\xi} J^i - 
	\partial_\xi s \partial_{\xi} j^i - 
	\partial_{\partial_{m}\xi}s \partial_{\xi} \partial_m j^i\right)  =\nonumber\\
=
\partial_i{\mathfrak{J}}^{i} + \partial_i(\partial_\xi s)j^i + \partial_i(\partial_{\partial_n\xi} s)\partial_n j^i \geq 0.
}
Then we can observe, that the last line of the above expression depends linearly on the fourth spatial derivative of the internal variable,  $\partial_{ijkl}\xi$. The coefficents of this term,  $\partial_{\partial_{ijk}\xi}\mathfrak{J}^i$ and $ \partial_i(\partial_{\partial_n\xi} s)\partial_{\partial_{ijk}\xi}j^i$, and also the remaining terms in the inequality are all independent on the fourth spatial derivative itself, they are only third order weakly nonlocal. Therefore the coefficient of  $\partial_{ijkl}\xi$ must be zero in the inequality. This is a repeated application of Liu procedure. Therefore:
\eqn{hmm}{
   \partial_{ijkl} \xi &: \partial_{\partial_{jkl}\xi} \mathfrak{J}^i =
		\partial_n(\partial_{\partial_{n}\xi}) \partial_{\partial_{jkl}\xi} j^n. 
}
The solution of this equation for the extra entropy flux is 
\eqn{eentfl}{
{\mathfrak{J}^i} = -\partial_n\left(\partial_{\partial_i \xi s}\right)j^n + \hat{\mathfrak{J}}^i(\xi,\partial_i\xi,\partial_{ij}\xi ). 
}
Here the remaining extra entropy flux, $\hat {\mathfrak{J}}^i$ is only second order weakly nonlocal. Therefore the final form of the entropy flux is
\eqn{CH_sflufin}{
 J^i=  \left(\partial_\xi s - \partial_n(\partial_{\partial_i\xi}  s) \right) j^n +
        (\partial_{\partial_n\xi}  s) \partial_n j^i+
        \hat{\mathfrak{J}}^{i}(\xi,\partial_i \xi,\partial_{ij}\xi ),
} 
This concludes a complete solution of the Liu system of equations, \re{liugl1}--\re{liugl2}. Considering all these conditions, and assuming a zero extra entropy flux, the dissipation  inequality reduces to the following simple expression:
\eqn{CH_eprfin}{
0 \leq \partial_i \big[\partial_\xi s - \partial_n( \partial_{\partial_n\xi}  s)\big]  j^i.
}

\re{CH_sflufin} and \re{CH_eprfin} are identical with the expressions \re{CH_sflu} and \re{CH_epr}, that we have obtained with the help of divergence separation method. 

The product form of the inequality, \re{CH_eprfin}, with the fourth order weakly nonlocal constitutive $j^i$ as a multiplier is a flux-force system, where the constitutive state space ensures a linear solution by Lagrange mean value theorem:
\eqn{CH_constt1}{
j^i =  -\kappa
	\partial^i\left(\partial_\xi \rs  -  \partial_k\left(\partial_{\partial_k\xi} \rs\right) \right),
} 
Substituting this expression into the balance \re{CH_bal1} we obtain the Cahn-Hilliard equation again:
\eqn{CH_balf1}{
\partial_t \xi - \partial_i \left[\kappa
	\partial^i\left(\partial_\xi \rs  -  \partial_k\left(\partial_{\partial_k\xi} \rs\right) \right)\right] =0.
}
The particular case of first order weakly nonlocal state space leads to the diffusive evolution of Classical Irreversible Thermodynamics \cite{Van02a}. This concludes the thermodynamic derivation.

\section{Discussion}

In this paper we have seen that phase-field evolution equations can be derived from the second law without variational considerations and functional derivatives. The intuitive method of separation of divergences enlights that the background is the separation of surface and bulk contributions. A more rigorous analysis with the application of Liu procedure led essentially to the same results when a constitutive entropy flux was applied with derivative constraints. This derivation of the evolution equations is based only on general principles and therefore the obtained Ginzburg-Landau and Cahn-Hilliard dynamics is universal. 

Let us emphasize that the differences between the pure thermodynamic approaches are not fundamental. They are negligible when compared to the difficulties of the doubled structures mentioned in the introduction. For example the simple and elegant approach of Heida, M\'alek and Rajagopal is based on the separation of divergences and the solution of the dissipation inequality. It is applied to obtain coupled diffusive thermal dynamics of fluid mixtures including Korteweg fluids \cite{HeiEta12a}. A more rigorous second law analysis, probably could lead to a justification of their results. An independent but similar work of Giorgi emphasizes the universality of the uniform thermodynamic derivation when compared to  the configurational force balance of Gurtin and the virtual power approach of Fremond \cite{Gio09a}.

On the other hand a less constructive but rigorous standpoint can lead to general but complicated conditions. These conditions are nevertheless useful to  improve a heuristic ansatz and to check its validity \cite{CimEta16a}. A similarly instructive analysis is the work of Pawlow, where the constitutive state space is first order weakly nonlocal in time, the entropy flux is constitutive but Pawlow does not apply a derivative prolongation of the constraints. Then the Liu equations cannot be solved in a closed function form, and to obtain the particular Cahn-Hilliard dynamics requires a further ad hoc restriction. However, the evolution equation of the Lagrange-Farkas multiplier leads to the microforce balance of Gurtin \cite{Paw06a}. That is, sacrificing the constructivity obscures the advantages of a uniform approach, but it still can be well interpreted.

Here the thermodynamic methods are demonstrated with scalar internal variables. The generalization for vectors and tensors is straightforward, as well as for classical thermodynamic variables and more complicated constraints (see e.g. \cite{VanFul06a,Van08a,Van09a1,VanPap10a}). We need to mention also, that time nonlocality, the treatment of memory and inertial effects is a different matter. For such systems dual internal variables are suitable for a uniform thermodynamic approach \cite{VanAta08a,BerEta11a,VanEta14a,BerVan17b}.


\section{Acknowledgement}   
The work was supported by the grants National Research, Development and Innovation Office – NKFIH​116197 (116375), NKFIH 124366 (124366), 123815 and 116197. 

This paper is dedicated to the memory of Gerard Maugin, to the scholar with wide interest and broad understanding, to the faithful champion of nonequilibrium thermodynamics.

\bibliographystyle{unsrt}

\begin{thebibliography}{10}

\bibitem{Gur96a}
M.~G. Gurtin.
\newblock Generalized {G}inzburg-{L}andau and {C}ahn-{H}illiard equations based
  on a microforce balance.
\newblock {\em Physica D}, 92:178--192, 1996.

\bibitem{Gio09a}
C.~Giorgi.
\newblock Continuum thermodynamics and phase-field models.
\newblock {\em Milan Journal of Mathematics}, 77(1):67--100, 2009.

\bibitem{Van13p2}
P.~V\'an.
\newblock Thermodynamics of continua: The challenge of universality.
\newblock In M.~Pilotelli and G.~P. Beretta, editors, {\em Proceedings of the
  12th Joint European Thermodynamics Conference}, 228--233, Brescia,
  2013. Cartolibreria SNOOPY, arXiv:1305.3582.

\bibitem{HohHal77a}
P.~C. Hohenberg and B.~I. Halperin.
\newblock Theory of dynamic critical phenomena.
\newblock {\em Reviews of Modern Physics}, 49(3):435--479, 1977.

\bibitem{HohKre15a}
P.~C. Hohenberg and A.P. Krekhov.
\newblock An introduction to the {G}inzburg-{L}andau theory of phase
  transitions and nonequilibrium patterns.
\newblock {\em Physics Reports}, 572:1--42, 2015.

\bibitem{LanGin50a}
L.~D. Landau and V.~L. Ginzburg.
\newblock K teorii sverkhrovodimosti.
\newblock {\em Zhurnal Eksperimentalnoi i Teoreticheskoi Fiziki}, 20:1064,
  1950.
\newblock In English: On the theory of superconductivity, in:
  Collected papers of L. D. Landau, ed. D. ter Haar, (Pergamon, Oxford, 1965),
  pp. 546-568.

\bibitem{LanKha54a}
L.~D. Landau and I.~M. Khalatnikov.
\newblock Ob anomal'nom pogloshchenii zvuka vblizi tochek fazovogo perekhoda
  vtorogo roda.
\newblock {\em Dokladu Akademii Nauk, SSSR}, 96:469--472, 1954.
\newblock English translation: On the anomalous absorption of sound near a
  second order transition point. in: Collected papers of L. D. Landau, ed. D.
  ter Haar,(Pergamon, Oxford, 1965), pp. 626-633.

\bibitem{Cah61a}
J.~W. Cahn.
\newblock On spinodal decomposition.
\newblock {\em Acta Metallica}, 9:795--801, 1961.

\bibitem{AllCah79a}
S.~M. Allen and J.~W. Cahn.
\newblock A microscopic theory for antiphase boundary motion and its
  application to antiphase domain coarsening.
\newblock {\em Acta Metallurgica}, 27(6):1085--1095, 1979.

\bibitem{CahHil58a}
J.~W. Cahn and J.~E. Hilliard.
\newblock Free energy of a nonuniform system {I}. {I}nterfacial free energy.
\newblock {\em Journal of Chemical Physics}, 28:258--267, 1958.

\bibitem{PenFif90a}
O.~Penrose and P.~C. Fife.
\newblock Thermodynamically consistent models of phase-field type for the
  kinetics of phase transitions.
\newblock {\em Physica D}, 43:44--62, 1990.

\bibitem{AltPaw92a}
H.W Alt and I.~Pawlow.
\newblock A mathematical model of dynamics of non-isothermal phase separation.
\newblock {\em Physica D: Nonlinear Phenomena}, 59(4):389--416, 1992.

\bibitem{Ant95a}
L.~K. Antanovskii.
\newblock A phase field model of capillarity.
\newblock {\em Physics of fluids}, 7(4):747--753, 1995.

\bibitem{Ant96a}
L.~K. Antanovskii.
\newblock Microscale theory of surface tension.
\newblock {\em Physical Review E}, 54(6):6285, 1996.

\bibitem{Grm08a}
M.~Grmela.
\newblock Extensions of classical hydrodynamics.
\newblock {\em Journal of Statistical Physics}, 132(3):581--602, 2008.

\bibitem{AndWei11a}
D.~Anders and K.~Weinberg.
\newblock A variational approach to the decomposition of unstable viscous
  fluids and its consistent numerical approximation.
\newblock {\em ZAMM-Zeitschrift f{\"u}r Angewandte Mathematik und Mechanik}, 91(8):609--629, 2011.

\bibitem{FabEta06a}
M.~Fabrizio, C.~Giorgi, and A.~Morro.
\newblock A thermodynamic approach to non-isothermal phase-field evolution in
  continuum physics.
\newblock {\em Physica D: Nonlinear Phenomena}, 214(2):144--156, 2006.

\bibitem{AndEta98a}
D.~M. Anderson, G.~B. McFadden, and A.~A. Wheeler.
\newblock Diffuse-interface methods in fluid mechanics.
\newblock {\em Annual Rev. in Fluid Mechanics}, 30:139--65, 1998.

\bibitem{Ott05b}
H.~C. \"Ottinger.
\newblock {\em Beyond equilibrium thermodynamics}.
\newblock Wiley-Interscience, 2005.

\bibitem{GrmOtt97a}
M.~Grmela and H.~C. \"Ottinger.
\newblock Dynamics and thermodynamics of complex fluids. {I}. {D}evelopment of
  a general formalism.
\newblock {\em Physical Review E}, 56(6):6620--6632, 1997.

\bibitem{OttGrm97a}
H.~C. \"Ottinger and M.~Grmela.
\newblock Dynamics and thermodynamics of complex fluids. {II}. {I}llustrations
  of a general formalism.
\newblock {\em Physical Review E}, 56(6):6633--6655, 1997.

\bibitem{PriStr95b}
I. Prigogine and I. Stengers. 
\newblock La nouvelle alliance, Gallimard, Paris 1986.

\bibitem{YouMan99b}
W.~Yourgrau and S.~Mandelstam.
\newblock {\em Variational principles in dynamics and quantum theory}.
\newblock Pitman, New York-Toronto-London, 2 edition, 1999.

\bibitem{VanMus95a}
P.~V\'an and W.~Muschik.
\newblock Structure of variational principles in nonequilibrium thermodynamics.
\newblock {\em Physical Review E}, 52(4):3584--3590, 1995.

\bibitem{VanNyi99a}
P.~V\'an and B.~Ny\'\i{}ri.
\newblock Hamilton formalism and variational principle construction.
\newblock {\em Annalen der Physik (Leipzig)}, 8:331--354, 1999.

\bibitem{Gya70b}
I.~Gyarmati.
\newblock {\em Non-equilibrium Thermodynamics /{F}ield Theory and Variational
  Principles/}.
\newblock Springer Verlag, Berlin, 1970.

\bibitem{SieFar05b}
S.~Sieniutycz and Farkas H., editors.
\newblock {\em Variational and Extremum Principles in Macroscopic Problems}.
\newblock Elsevier, Amsterdam-etc., 2005.

\bibitem{Ver14a}
J.~Verh\'as.
\newblock Gyarmati's variational principle of dissipative processes.
\newblock {\em Entropy}, 16:2362--2383, 2014.

\bibitem{Gla15a}
K.S.~Glavatskiy.
\newblock Lagrangian formulation of irreversible thermodynamics and the second
  law of thermodynamics.
\newblock {\em The Journal of Chemical Physics}, 142(20):204106, 2015.

\bibitem{MatAta05a}
T.~Matolcsi, P.~V\'an, and Verh\'as J.
\newblock Fundamental problems of variational principles: objectivity,
  symmetries and construction.
\newblock In S.~Sieniutycz and Farkas H., editors, {\em Variational and
  Extremum Principles in Macroscopic Problems}, 57--74. Elsevier,
  Amsterdam-etc., 2005.

\bibitem{Cap85a}
G.~Capriz.
\newblock Continua with latent microstructure.
\newblock {\em Archive for Rational Mechanics and Analysis}, 90(1):43--56,
  1985.

\bibitem{Cap89b}
G.~Capriz.
\newblock {\em Continua with Microstructure}.
\newblock Springer Verlag, New York-etc., 1989.

\bibitem{Mar02a}
P.~M. Mariano.
\newblock Multifield theories in mechanics of solids.
\newblock {\em Advances in Applied Mechanics}, 38:1--94, 2002.

\bibitem{Gro59b}
S.R. de~Groot.
\newblock {\em Thermodynamics of irreversible processes}.
\newblock North Holland, 1959.

\bibitem{GroMaz62b}
S.~R. de~Groot and P.~Mazur.
\newblock {\em Non-equilibrium Thermodynamics}.
\newblock North-Holland Publishing Company, Amsterdam, 1962.

\bibitem{Mau99b}
G.~Maugin.
\newblock {\em The thermomechanics of nonlinear irreversible behaviors ({A}n
  introduction)}.
\newblock World Scientific, Singapore-New Jersey-London-Hong Kong, 1999.

\bibitem{HeiEta12a}
M.~Heida, J.~M{\'a}lek, and K.R. Rajagopal.
\newblock On the development and generalizations of cahn--hilliard equations
  within a thermodynamic framework.
\newblock {\em Zeitschrift f{\"u}r Angewandte Mathematik und Physik},
  63(1):145--169, 2012.

\bibitem{ColNol63a}
B.~D. Coleman and W.~Noll.
\newblock The thermodynamics of elastic materials with heat conduction and
  viscosity.
\newblock {\em Archive for Rational Mechanics and Analysis}, 13:167--178, 1963.

\bibitem{Liu72a}
I-Shih Liu.
\newblock Method of {L}agrange multipliers for exploitation of the entropy
  principle.
\newblock {\em Archive of Rational Mechanics and Analysis}, 46:131--148, 1972.

\bibitem{ColMiz67a}
B.~D. Coleman and V.~J. Mizel.
\newblock Existence of entropy as a consequence of asymptotic stability.
\newblock {\em Archive for Rational Mechanics and Analysis}, 25:243--270, 1967.

\bibitem{MusEhr96a}
W.~Muschik and H.~Ehrentraut.
\newblock An amendment to the {S}econd {L}aw.
\newblock {\em Journal of Non-Equilibrium Thermodynamics}, 21:175--192, 1996.

\bibitem{TriAta08a}
V.~Triani, C.~Papenfuss, V.~A. Cimmelli, and W.~Muschik.
\newblock Exploitation of the {S}econd {L}aw: {C}oleman-{N}oll and {L}iu
  procedure in comparison.
\newblock {\em Journal of Non-Equilibrium Thermodynamics}, 33:47--60, 2008.

\bibitem{Van17a}
P.~V\'an.
\newblock Galilean relativistic fluid mechanics.
\newblock {\em Continuum Mechanics and Thermodynamics}, 29(2):585--610, 2017.
\newblock arXiv:1508.00121.

\bibitem{ColGur67a}
B.~D. Coleman and M.~E. Gurtin.
\newblock Thermodynamics with internal state variables.
\newblock {\em The Journal of Chemical Physics}, 47(2):597--613, 1967.

\bibitem{Gur65a}
M.~E. Gurtin.
\newblock Thermodynamics and the possibility of spatial interaction in elastic
  materials.
\newblock {\em Archive for Rational Mechanics and Analysis}, 19:339--352, 1965.

\bibitem{DunSer85a}
J.~E. Dunn and J.~Serrin.
\newblock On the thermomechanics of interstitial working.
\newblock {\em Archive of Rational Mechanics and Analysis}, 88:95--133, 1985.

\bibitem{Gur00b}
M.~E. Gurtin.
\newblock {\em Configurational Forces as Basic Concepts of Continuum Physics}.
\newblock Springer, New York-etc., 2000.

\bibitem{Ger73a}
P.~Germain.
\newblock The method of virtual power in continuum mechanics. {P}art 2:
  Microstructure.
\newblock {\em SIAM Journal of Applied Mathematics}, 25:556--575, 1973.

\bibitem{Mau80a}
G.~A. Maugin.
\newblock The principle of virtual power in continuum mechanics. {A}pplication
  to coupled fields.
\newblock {\em Acta Mechanica}, 35:1--70, 1980.

\bibitem{Fre01b}
M.~Fr{\'e}mond.
\newblock {\em Non-smooth Thermomechanics}.
\newblock Springer, 2001.

\bibitem{Mau13a}
G.A. Maugin.
\newblock The principle of virtual power: from eliminating metaphysical forces
  to providing an efficient modelling tool.
\newblock {\em Continuum Mechanics and Thermodynamics}, 25:127--146, 2013.

\bibitem{Paw06a}
I.~Paw\l ow.
\newblock Thermodynamically consistent {C}ahn-{H}illiard and {A}llen-{C}ahn
  models in elastic solids.
\newblock {\em Discrete and Continuous Dynamical Systems}, 15(4):1169--1191,
  2006.

\bibitem{CimEta16a}
V.A.~Cimmelli, F.~Oliveri, and A.R.~Pace.
\newblock Phase-field evolution in {Cahn--Hilliard--Korteweg} fluids.
\newblock {\em Acta Mechanica}, 227(8):2111--2124, 2016.

\bibitem{Mau06a}
G.~A. Maugin.
\newblock On the thermomechanics of continuous media with diffusion and/or weak
  nonlocality.
\newblock {\em Archive of Applied Mechanics}, 75:723--738, 2006.

\bibitem{Pen04b}
R.~Penrose.
\newblock {\em The Road to Reality}.
\newblock Jonathan Cape, 2004.

\bibitem{MauDro83a}
G.~A. Maugin and R.~Drouot.
\newblock Internal variables and the thermodynamics of macromolecule solutions.
\newblock {\em International Journal of Engineering Science}, 21(7):705--724,
  1983.

\bibitem{BerVan17b}
A.~Berezovski and V\'an P.
\newblock {\em Internal Variables in Thermoelasticity}.
\newblock Springer, 2017.

\bibitem{Van08a}
P.~V\'an.
\newblock Internal energy in dissipative relativistic fluids.
\newblock {\em Journal of Mechanics of Materials and Structures},
  3(6):1161--1169, 2008.
\newblock arXiv:07121437 [nucl-th].

\bibitem{BedAta03a}
D.~Bedeaux, E.~Johannessen, and A.~Rojorde.
\newblock A nonequilibrium van der {W}aals square gradient model. ({I}). the
  model and its numerical solution.
\newblock {\em Physica A}, 330:329--353, 2003.

\bibitem{JohBed03a}
E.~Johannessen and D.~Bedeaux.
\newblock A nonequilibrium van der {W}aals square gradient model. ({II}). local
  equilibrium of the {G}ibbs surface.
\newblock {\em Physica A}, 330:354--372, 2003.

\bibitem{JohBed04a}
E.~Johannessen and D.~Bedeaux.
\newblock A nonequilibrium van der {W}aals square gradient model. ({III}). heat
  and mass transfer coefficients.
\newblock {\em Physica A}, 336:252--270, 2004.

\bibitem{GlaBed08a}
K.~S. Glavatskiy and D.~Bedeaux.
\newblock Nonequilibrium properties of a two-dimensionally isotropic interface
  in a two-phase mixture as described by the square gradient model.
\newblock {\em Physical Review E}, 77:061101, 2008.

\bibitem{Cim07a}
V.~A. Cimmelli.
\newblock An extension of {L}iu procedure in weakly nonlocal thermodynamics.
\newblock {\em Journal of Mathematical Physics}, 48:113510, 2007.

\bibitem{Van02a}
P.~V\'an.
\newblock Weakly nonlocal irreversible thermodynamics - the {G}inzburg
  -{L}andau equation.
\newblock {\em Technische Mechanik}, 22(2):104--110, 2002.
\newblock (cond-mat/0111307).

\bibitem{VanFul06a}
P.~V\'an and T.~F\"ul\"op.
\newblock Weakly nonlocal fluid mechanics - the {S}chr\"odinger equation.
\newblock {\em Proceedings of the Royal Society, London A}, 462(2066):541--557,
  2006.
\newblock (quant-ph/0304062).

\bibitem{Van09a1}
P.~V\'an.
\newblock Weakly nonlocal non-equilibrium thermodynamics - variational
  principles and {S}econd {L}aw.
\newblock In Ewald Quak and Tarmo Soomere, editors, {\em Applied Wave
  Mathematics (Selected Topics in Solids, Fluids, and Mathematical Methods)},
  chapter III, pages 153--186. Springer-Verlag, Berlin-Heidelberg, 2009.
\newblock (arXiv:0902.3261).

\bibitem{VanPap10a}
P.~V\'an and C.~Papenfuss.
\newblock Thermodynamic consistency of third grade finite strain elasticity.
\newblock {\em Proceedings of the Estonian Academy of Sciences},
  59(2):126--132, 2010, arXiv:1002.2965v1.

\bibitem{VanAta08a}
P.~V\'an, A.~Berezovski, and J.~Engelbrecht.
\newblock Internal variables and dynamic degrees of freedom.
\newblock {\em Journal of Non-Equilibrium Thermodynamics}, 33(3):235--254,
  2008.
\newblock cond-mat/0612491.

\bibitem{BerEta11a}
A.~Berezovski, J.~Engelbrecht, and G.~A. Maugin.
\newblock Generalized thermomechanics with dual internal variables.
\newblock {\em Archive of Applied Mechanics}, 81(2):229--240, 2011.

\bibitem{VanEta14a}
P.~V\'an, C.~Papenfuss, and A.~Berezovski.
\newblock Thermodynamic approach to generalized continua.
\newblock {\em Continuum Mechanics and Thermodynamics}, 25(3):403--420, 2014.
\newblock Erratum: 421-422, arXiv:1304.4977.

\end{thebibliography}

\end{document}